\documentclass[aps,preprint,showpacs]{revtex4}
\begin{document}           
\draft
\input{psfig.tex}
\title{Transport induced by Density Waves \\ in a Andreev-Lifshitz Supersolid}
\author{Kwang-Hua W. Chu \cite{2007:CHU}}
\affiliation{P.O. Box 30-15, Shanghai 200030, PR China}
\begin{abstract}
Macroscopic derivation of the
entrainment in in a Andreev-Lifshitz Supersolid induced by a surface elastic wave
propagating along the flexible  interface is
conducted by considering the nonlinear coupling between the
interface and the rarefaction effect.
We obtain the critical bounds for  zero-volume-flow-rate states
corresponding to specific rarefaction measure and wave number which is
relevant to
the rather small critical velocity of supersolid flows reported by
Kim and Chan.\newline

\end{abstract}
\maketitle                 
\bibliographystyle{plain}
In 1969, it was conjectured by Andreev and Lifshitz$^{1}$
that at zero temperature, delocalized defects may exist
in a quantum solid, as a result of which the number of
sites of an ideal crystal lattice may not coincide with the
total number of particles. Originally, this conjecture was
proposed for three dimensional quantum solids made of
atoms ($^3$He, $^4$He, $\cdots$) which do not interact via Coulomb
repulsion.
The proposed {\it supersolid} phase is believed to occur due to the quantum
behavior of point defects, namely vacancies and interstitials, in this
crystal of bosons$^{2-3}$.
Researchers have found that
a small lattice model does not exhibit the mesoscopic signature
of an intermediate phase separating the solid from
the liquid, where the solid and the fluid would coexist$^{4}$.
Such a vacancy-solid phase was indeed suggested$^{1}$ by
Andreev and Lifshitz if the zero point motions of certain
defects become sufficient to form waves propagating inside
the solid.  \newline
Castaing and Nozi\`{e}res have later considered$^{5}$
such a possibility for spin polarized $^3$He. The statistics of
the defects depend on their nature.
For simple vacancies
in the crystal, their statistics is given by the statistics of
the particles out of which the solid is made. If the defects
are bosons, they may form a condensate, giving rise to a
superfluid coexisting with the solid. This supersolid phase
is discussed in certain bosonic models$^{6}$. If the defects
are fermions, they may form a Fermi liquid$^{7}$ coexisting
with the solid, such that the system is neither a solid,
nor a liquid.
Two kinds of motion are possible in it; one
possesses the properties of motion in an elastic solid, the
second possesses the properties of motion in a liquid.
This interesting issue motivates our present study.  \newline
Early theoretical work by Andreev and Lifshitz$^{1}$ and Chester$^{2}$
showed that solids may feature a Bose-Einstein condensate
of vacancies (or interstitial atoms) and thus possess
superfluid (SF) properties. Quite recently one description of the quantum
solid is as a density wave that has formed in the
quantum fluid$^{8-9}$. The periodicity of this density wave need not match
precisely to the particle density, so that the ground state may be
incommensurate, with unequal densities of
vacancies and interstitials. Whether or not the
same is true for quantum {\it fluctuations} is not clear at this point.
We noticed that previous theories imply a corresponding
vacancy contribution to the specific heat that is as large as the
phonon contribution near
1 Kelvin$^{3,10}$.
Based on these considerations or
phenomenological approaches, in this letter, we shall demonstrate
that  wavy flexible interface (between atoms and free vacancies
or defects) or highly-pressured environments$^{8}$
can produce elastically deformed interface or peristaltic motion will
induce time-averaged transport in a Andreev-Lifshitz supersolid.
\newline
Theoretical
studies of interphase nonlocal transport phenomena which appear as
a result of a different type of
nonequilibrium representing propagation of a surface elastic wave
have been performed before$^{11-12}$.
These are relevant to particles  flowing along deformable elastic
slabs with the dominated
parameter being the Knudsen number (Kn = mean-free-path/$L_d$,
mean-free-path (mfp) is the mean free path of the particles, $L_d$ is
proportional to
the
distance between two slabs)$^{13-15}$. The role of the Knudsen number is
similar to that of
the Navier slip parameter $N_s (= \mu S/L_d$;
 S is a proportionality constant as $u_s = S \tau$,
$\tau$ :
the shear stress of the bulk velocity; $u_s$ : the dimensional slip velocity;
for a
no-slip case, $S = 0$, but for a no-stress condition. $S=\infty$, $\mu$ is the
 viscosity). \newline
We shall choose a periodic domain
to simplify
our mathematical treatments. The flat interface is presumed.
We adopt the macroscopic or hydrodynamical approach and simplify the original
system of equations (related to the momentum and mass transport)
to one single higher-order quasi-linear partial differential
equation in terms of the unknown stream function. In this study,
as the temperature is rather low and the phase is related to the supersolid
(there might be weakly friction or shearing dissipation in-between)
we shall assume that
 the governing
equations are the incompressible Navier-Stokes equations which will be
associated with the microscopically slip velocity boundary conditions
along the interfaces$^{13-15}$. 
To consider the originally quiescent environment for
simplicity, due to the difficulty in solving a fourth-order
quasi-linear complex ordinary differential equation (when the wavy
boundary condition are imposed), we can finally get an
analytically perturbed solution and calculate those physical
quantities, like, time-averaged transport or entrainment,
 critical  forcing corresponding to
the freezed or zero-volume-flow-rate states. \newline
We consider a two-dimensional region of uniform thickness.
The flat-plane
boundaries   or   interfaces
are rather flexible and presumed to be elastic, on which are imposed
traveling sinusoidal waves of small amplitude $a$ (possibly due to
quantum fluctuations).
The vertical displacements of the upper and lower interfaces ($y=L_d$ and
$-L_d$) are thus presumed to be $\eta$ and $-\eta$, respectively,
where $\eta=a \cos [2\pi (x-ct)/\lambda$], $\lambda$ is the
wave length, and $c$ the wave speed. $x$ and $y$ are Cartesian
coordinates, with $x$ measured in the direction of wave
propagation and $y$ measured in the direction normal to the mean
position of the  interfaces. 
We have a characteristic
velocity $c$ and three characteristic lengths $a$, $\lambda$, and
$L_d$. The following variables based on $c$ and $L_d$ could thus be
introduced :
 $x'={x}/{L_d}$, 
 $y'={y}/{L_d}$, 
 $u'={u}/{c}$, 
 $v'={v}/{c}$, 
 $\eta'={\eta}/{L_d}$, 
 $\psi'={\psi}/({c\,L_d})$, 
 $t'={c\,t}/{h}$, 
 $p'={p}/({\rho c^2})$,
where $\psi$ is the dimensional stream function, $u$ and $v$ are the velocities
along the $x$- and $y$-directions; $\rho$ is the density, $p$
is  the pressure. The primes could be dropped in the following. The amplitude
ratio $\epsilon$, the wave number $\alpha$, and the Reynolds
number $Re$ (representing the weakly friction or shearing dissipation
using a viscosity $\nu$) are defined by
 $\epsilon={a}/{L_d}$, 
 $\alpha={2 \pi L_d}/{\lambda}$,  
 $Re ={c\,L_d}/{\nu}$.
We shall seek a solution in the form of a series in the small parameter
$\epsilon$ :
 $\psi=\psi_0 +\epsilon \psi_1 + \epsilon^2 \psi_2 + \cdots$,
 ${\partial p}/{\partial x}=({\partial p}/{\partial
 x})_0+\epsilon ({\partial p}/{\partial x})_1 +\epsilon^2 ({\partial
 p}/{\partial x})_2 +\cdots$,
with $u=\partial \psi/\partial y$, $v=-\partial \psi/\partial x$.
The 2D (x- and y-) momentum equations and the equation of
continuity$^{16-17}$ could be in terms of the stream function $\psi$ if the
$p$-term  is eliminated. The final governing equation is
\begin{equation}
 \frac{\partial}{\partial t} \nabla^2 \psi + \psi_y \nabla^2 \psi_x
 -\psi_x \nabla^2 \psi_y =\frac{1}{Re}\nabla^4 \psi,
\hspace*{12mm} \nabla^2 \equiv\frac{\partial^2}{\partial x^2}
+\frac{\partial^2}{\partial y^2}   ,
\end{equation}
and subscripts indicate the partial differentiation. Thus, we have
\begin{equation}
 \frac{\partial}{\partial t} \nabla^2 \psi_0 +\psi_{0y} \nabla^2
 \psi_{0x}-\psi_{0x} \nabla^2 \psi_{0y}=\frac{1}{Re} \nabla^4 \psi_0
 ,
\end{equation}
\begin{equation}
 \frac{\partial}{\partial t} \nabla^2 \psi_1 +\psi_{0y} \nabla^2
 \psi_{1x}+\psi_{1y}\nabla^2 \psi_{0x}-\psi_{0x} \nabla^2 \psi_{1y}-
 \psi_{1x} \nabla^2 \psi_{0y}=\frac{1}{Re} \nabla^4 \psi_1
 ,
\end{equation}
and other higher order terms. Microscopic boundary
conditions imposed by the symmetric motion of the  interfaces and the
non-zero slip velocities$^{13-14}$ are : $u=\mp$ Kn $\,du/dy$, $v=\pm
\partial \eta/\partial t$ at $y=\pm (1+ \eta)$, here Kn=mfp$/L_d$.
The boundary conditions are expanded in powers of $\eta$ and
then $\epsilon$.
These equations, together with the condition of symmetry and a
uniform  $(\partial
p/\partial x)_0$, yield :
 $\psi_0 =K_0 [ (1+2 \mbox{Kn}) y-{y^3}/{3}]$, 
 $K_0={Re}(-{\partial p}/{\partial x})_0 /2$, 
 $\psi_1 =\{\phi(y) e^{i \alpha (x-t)}+\phi^* (y) e^{-i \alpha
 (x-t)} \}/{2}$ , 
where the asterisk denotes the complex conjugate. A substitution
of $\psi_1$ into Eqn. (3) yields
\begin{displaymath}
 \{\frac{d^2}{d y^2} -\alpha^2 +i \alpha Re [1-K_0 (1-y^2+2 \mbox{Kn})]\}
 (\frac{d^2}{d y^2} -\alpha^2) \phi -2 i\alpha K_0 Re \,\phi =0  
\end{displaymath}
or
if originally the environment is quiescent : $K_0 = 0$ (this
corresponds to a free pumping case)
 $({d^2}/{d y^2} -\alpha^2) ({d^2}/{d y^2} -\bar{\alpha}^2) \phi
 =0$,   
 $\bar{\alpha}^2= \alpha^2 -i \alpha Re$. 
The boundary conditions are
 $\phi_y (\pm 1) \pm\phi_{yy} (\pm 1) \mbox{Kn}=0$,
 $\phi (\pm 1)=\pm 1 .$  
Similarly, with
 $ \psi_2= \{D(y)+E(y) e^{i 2\alpha (x-t)} +E^* (y) e^{-i
  2\alpha (x-t)} \}/2$ ,  
we have much more complicated equations (cf. Chu in [17])
and the boundary conditions
\begin{equation}
 D_y (\pm 1) +\frac{1}{2} [\phi_{yy} (\pm 1)+\phi^*_{yy} (\pm 1)] =
 \mp \mbox{Kn}  \{\frac{1}{2} [\phi_{yyy} (\pm 1) +
 \phi^*_{yyy} (\pm 1)]+ D_{yy} (\pm 1) \},  
\end{equation}
 $E_y (\pm 1)+\phi_{yy} (\pm 1)/2 
  =\mp \mbox{Kn} [\phi_{yyy} (\pm 1)/2 + E_{yy} (\pm 1) ]$,
 $E(\pm 1)+ \phi_y  (\pm 1)/4 =0$.   
After lengthy algebraic manipulations, we obtain
 $\phi=c_0 e^{\alpha y}+c_1 e^{-\alpha y}+c_2 e^{\bar{\alpha} y}+
      c_3 e^{-\bar{\alpha} y}$,
where $c_i, i=0,1,2,3$
are related to $\alpha$, $\bar{\alpha}$, and Kn
(cf. Chu in Ref. 17).
\newline To obtain a simple
solution which relates to the mean transport so long as only terms of
$O(\epsilon^2)$ are concerned, we see that if every term in the
x-momentum equation is averaged over an interval of time equal to
the period of oscillation, we obtain
\begin{equation}
 \overline{\frac{\partial p}{\partial x}}=\epsilon^2 \overline{(\frac{\partial
 p}{\partial x})_2} =\epsilon^2 [\frac{D_{yyy}}{2 Re} + \frac{i Re}{4}
 (\phi \phi^*_{yy} -\phi^* \phi_{yy})] +O (\epsilon^3) =\epsilon^2
 \frac{\Pi_0}{Re} +O(\epsilon^3) ,   
\end{equation}
where $\Pi_0$ is the integration constant
 and could be fixed indirectly in the coming equation
below. Now, from Eqn. (4), we have
\begin{equation}
  D_y (\pm 1) \pm \mbox{Kn} D_{yy} (\pm 1)= -\frac{1}{2} [\phi_{yy} (\pm 1)+
  \phi^*_{yy} (\pm 1)] \mp \mbox{Kn}  \{\frac{1}{2} [\phi_{yyy} (\pm 1)
+\phi^*_{yyy} (\pm 1)] \} ,  
\end{equation}
where $D_y (y)= \Pi_0 y^2 +a_1 y+ a_2 + {\cal C} (y)$, and together
from the expression of $\psi_2$, we can obtain ${\cal C} (y)$ which
also depends on the $\alpha$,  $\bar{\alpha}$, $Re$, $c_i, i=0,1,2,3$
(cf. Chu in Ref. 17).
In realistic applications we
must determine $\Pi_0$ from considerations of  outlet conditions
of the slab-region. $a_1$ equals to zero because of the symmetry of
boundary conditions.  \newline Once $\Pi_0$ is specified,
our solution for the mean speed ($u$ averaged over time) of superflow
is
 ${U}=\epsilon^2 {D_y}/{2}={\epsilon^2} \{ {\cal
 C}(y)-{\cal C} (1)+R_0-\mbox{Kn} \,{\cal C}_y (1)+\Pi_0
 [y^2-(1+2 \mbox{Kn})] \}/{2}$ 
where $R_0$ $=-\{ [\phi_{yy} (1)+\phi^*_{yy} (1)]$ $- \mbox{Kn}
[\phi_{yyy} (1)$ $+\phi^*_{yyy} (1)]\}/2$. To illustrate our results clearly, we adopt $U(Y) \equiv u(y$)
for the time-averaged results with $y\equiv Y$ in the following. \newline
Our numerical calculations confirm that the mean streamwise
velocity distribution (averaged over time) due to the induced
motion by the wavy elastic  interface in the case of free (vacuum) pumping
is dominated by $R_0$
(or Kn) and the parabolic distribution $-\Pi_0 (1-y^2)$. $R_0$
which defines the boundary value of $D_y$ has its origin in the
y-gradient of the first-order streamwise velocity distribution.
Note that the Reynolds number
here is based on the wave speed.  The physical trend herein is also
the same as those reported before$^{13-14,17}$ for the
slip-flow effects. The slip produces decoupling with the inertia of the wavy interface.
\newline Now, let us define a critical reflux condition as one for
which the mean velocity ${U} (Y)$ equals to zero at the
center-line $Y=0$. With the equation of $U$, we have
 $ \Pi_{0_{cr}}=Re \overline{({\partial p}/{\partial x})_2}=-{[{\alpha^2
  Re^2}F(0)/200 +\mbox{Kn} \,{\cal C}' (1)-R_0]}/{(1+2 \mbox{Kn})}$
which means the critical reflux condition is reached when $\Pi_0$
has above value. Pumping against a positive  forcing
greater than the critical value would result in a backward transport
(reflux) in the central region of the stream. This critical value
depends on $\alpha$, $Re$, and Kn. There will be no reflux if
the  pressure gradient is smaller than this $\Pi_0$.
Thus, for some $\Pi_0$ values less than  $\Pi_{0_{cr}}$, the superflow
will keep moving  forward. On the contrary, parts of the flow will move
 backward
if $\Pi_0 > \Pi_{0_{cr}}$. \newline
As reported in Ref. 8,  the rather small critical velocity ($\le 20 \mu$m/s) observed
shows an apparent dissipation or attenuation of the superflow. Thus,
we
present some of the values of $\Pi_0(\alpha, Re; \mbox{Kn}=0, 0.1)$ corresponding to
freezed or zero-volume-flow-rate states ($\int_{-1}^1 U(Y) dY=0$) in  Table 1 where
the wave number ($\alpha$) has the range between $0.10$ and $0.90$;
the Reynolds number ($Re$)$=0.1,1,10,50,100$.  We observe that as Kn increases from zero
to 0.1, the
critical $\Pi_0$  decreases
significantly (cf. Fig. 1). For the same Kn, once Re is larger than 10,
critical reflux values $\Pi_0$ drop rapidly and the wave-modulation
effect (due to $\alpha$) appears. The latter observation might be
interpreted as the strong coupling between the  interface and
the inertia of the streaming superflow. The illustration of the velocity
fields
for those zero-flux (zero-volume-flow-rate) or freezed states are shown
in Figure 2.
There are three wave numbers : $\alpha=0.2, 0.5, 0.8$. The Reynolds number
is 50.
Both no-slip and slip (Kn=0.1) cases are presented. The arrows
for slip cases are schematic
and represent the direction of positive and negative velocity fields. \newline
Some remarks could be
made about these states : the transport
being freezed in the time-averaged sense
for specific dissipations (in terms of Reynolds number which is the ratio of wave-inertia
and viscous shearing effects) and wave numbers
(due to the wavy interface or vacuum fluctuations)
for either no-slip and slip cases.
 Meanwhile, the time-averaged transport induced by the wavy interface
is proportional to the square
of the amplitude ratio (although the small amplitude waves being presumed), as can
be seen in Eqn. (5), which is qualitatively the
same as that presented in Ref. 11 for analogous interfacial problems.
\newline In brief summary,
the entrained transport (either postive or negative and there is possibility : freezing)
due to the wavy  interface
 is mainly tuned
by the  $\Pi_0$ for fixed Re. Meanwhile,
$\Pi_{0_{cr}}$ depends strongly on the Knudsen number
(Kn, a rarefaction measure) instead of Re or $\alpha$. These results
(cf. Table 1 and Fig. 1)
might
explain why there are rather small critical velocities
for superflows  in the temperature range where
a supersolidity is observed$^{8}$. Finally we like to stress that once the
phase is solid then the elastically tensile stress should be larger than
that of the
liquid phase and the latter could be a crucial test for the existence of the
supersolidity.

%
%
\newpage
%
\psfig{file=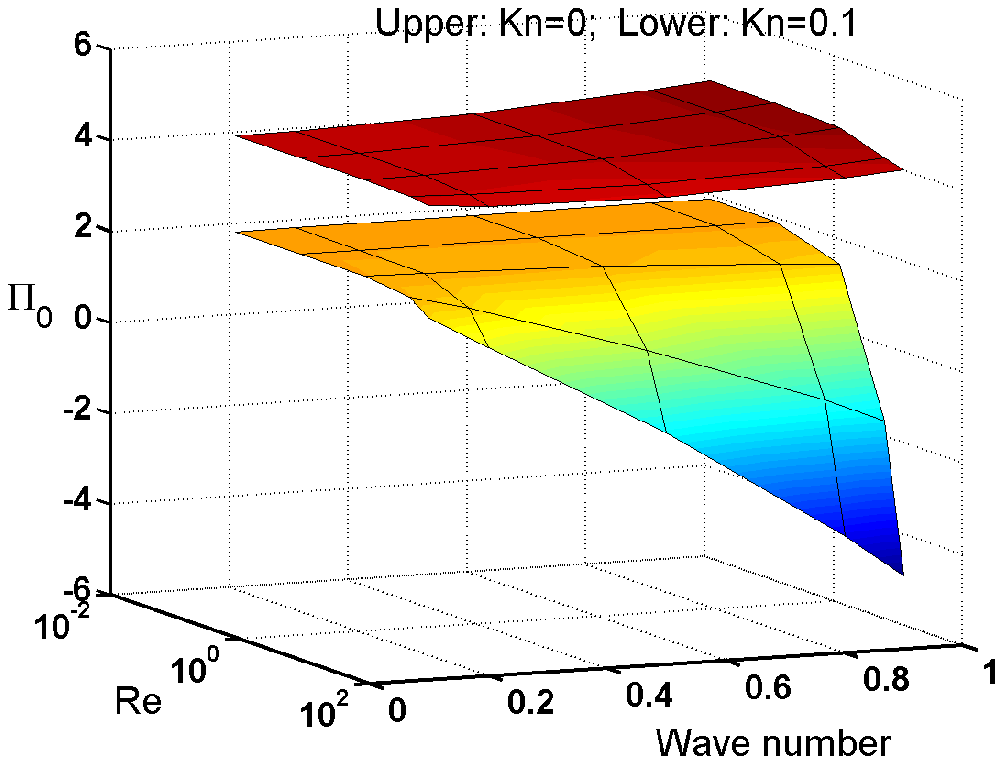,bbllx=0cm,bblly=13.8cm,bburx=12cm,bbury=24cm,rheight=10cm,rwidth=10cm,clip=}

\begin{figure} [h]
\hspace*{3mm} Fig. 1 \hspace*{1mm} Demonstration of Kn, Re and
$\alpha$ effects on the $\Pi_0$ (zero-flux states).
\newline \hspace*{3mm} 
$Re$ is the Reynolds number (the ratio of the wave-inertia and viscous
shearing dissipation).   \newline \hspace*{3mm}
$\alpha$ is the wave number and Kn is the Knudsen number
(a rarefaction measure).
\end{figure}

\newpage

\vspace{3mm}
\psfig{file=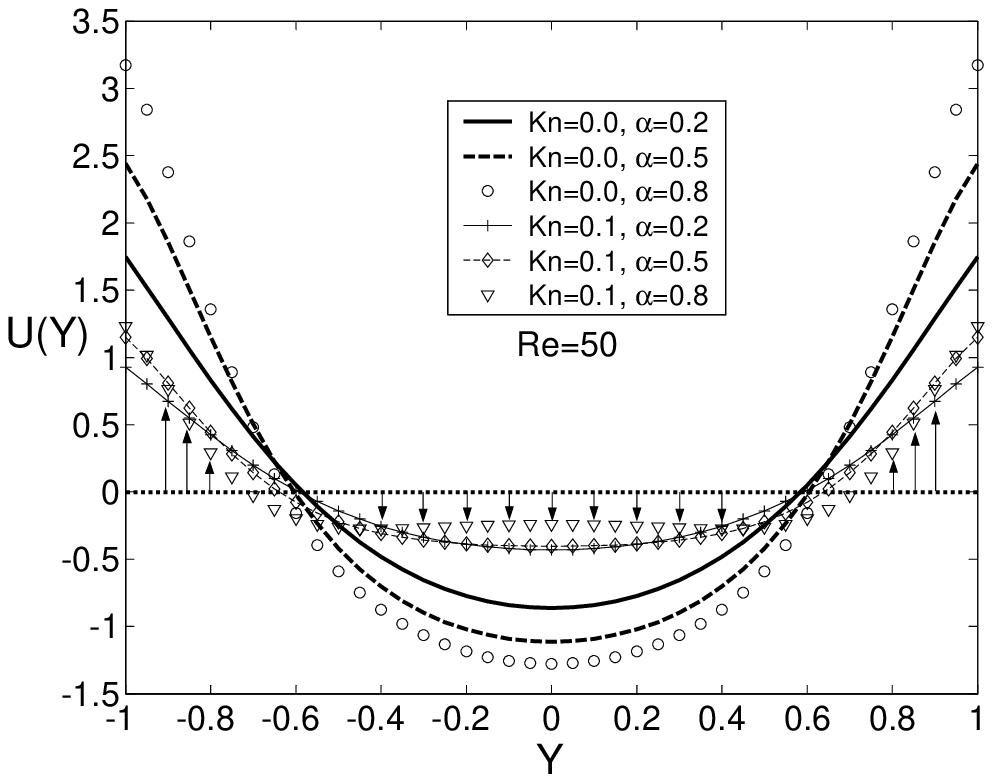,bbllx=0.0cm,bblly=12.5cm,bburx=12cm,bbury=24cm,rheight=9.8cm,rwidth=9.8cm,clip=}
%
\begin{figure} [h]
\hspace*{3mm} Fig. 2 \hspace*{1mm} Demonstration of the zero-flux states :
the mean velocity
field $U(Y)$ for \newline \hspace*{3mm}
wave numbers $\alpha=0.2,0.5,0.8$. The Reynolds number is $50$.
Kn is the rarefaction measure
\newline \hspace*{3mm}
(the mean free path of the particles divided by the characteristic length).
\newline \hspace*{3mm}
The arrows are schematic and illustrate the directions of positive and negative $U(Y)$.
\newline \hspace*{3mm}
The integration of $U(Y)$ w.r.t. $Y$ for these velocity fields gives zero volume flow rate.
\end{figure}

%
\newpage

\begin{table}[h]
 \caption{Zero-flux or freezed states values ($\Pi_0$) for a flat
 wavy interface.}
\vspace*{5mm}
\begin{center}
\begin{tabular}[b]{|r|c|c|c|c|c|c|}      \hline
    &          &  Re  &   &     &     &      \\ \hline
 Kn & $\alpha$ &  0.1 & 1 & 10 & 50 & 100 \\ \hline
 0  & 0.1  &  4.5088  & 4.5087  &   4.5078       &      4.4863 &         4.4316 \\  \cline{2-7}
    & 0.2  &  4.5269  & 4.5269  &   4.5231       &      4.4496 &         4.3275  \\  \cline{2-7}
    & 0.5  &  4.6586  & 4.6584  &   4.6359       &      4.4086 &         4.2682  \\  \cline{2-7}
    & 0.8  &   4.9238 &  4.9234 &    4.8708      &      4.5714 &         4.4488   \\ \cline{2-7}
    & 0.9  &  5.0475  &  5.0468 &    4.9827      &      4.6709 &        4.5516    \\ \hline
0.1  & 0.1 &  2.3983  & 2.3982  &   2.3927       &      2.2676 &        1.9560  \\ \cline{2-7}
     & 0.2 &  2.4003  & 2.4000  &   2.3774       &      1.9532 &        1.2217    \\ \cline{2-7}
     & 0.5 &  2.4149  & 2.4132  &   2.2731       &      0.7728 &       -0.9054 \\ \cline{2-7}
     & 0.8 &  2.4422  & 2.4379  &   2.0718       &     -0.5885 &      -3.4151  \\ \cline{2-7}
     & 0.9 &  2.4542  & 2.4486  &   1.9825       &     -1.1191 &        -4.3972  \\ \hline
\end{tabular}            
\end{center}
\end{table}
\end{document}